\documentclass[preprint,review,12pt]{elsarticle}    

\usepackage{amsmath,amsfonts}
\usepackage{url}

\usepackage{comment}   
\usepackage{booktabs}  
\usepackage{gensymb}   
\usepackage[figuresright]{rotating} 

\newcommand{\bi}{\begin{itemize}}
\newcommand{\ei}{\end{itemize}}

\def\be#1\ee{\begin{equation}#1\end{equation}}
\def\barr#1\earr{\begin{align}#1\end{align}}

\includecomment{review}
\renewcommand{\comment}[2]{}
\begin{review}
\renewcommand{\comment}[2]{\medskip\noindent\textcolor{#1}{\footnotesize #2}\medskip}
\end{review}

\begin{document}

\begin{frontmatter}

\title{
Visible light backscattering with applications to
the Internet of Things: State-of-the-art, 
challenges, and opportunities\tnoteref{t1}}

\tnotetext[t1]{An earlier version of this paper \cite{Ullah2022}
was presented at the 1st International
Workshop on Mobile and Pervasive Sensing
for Healthcare (MobiSens4H),
August 9-11, 2022, Niagara Falls, Ontario, Canada,
pp. 745--752
[DOI: 10.1016/j.procs.2022.07.111].}

\author[unina]{Muhammad Habib Ullah}
\ead{muhammad.habibullah@unina.it}

\author[unina,cnit]{Giacinto~Gelli}
\ead{gelli@unina.it}

\author[unina,cnit]{Francesco~Verde\corref{cor1}}
\ead{f.verde@unina.it}
        
 \cortext[cor1]{Corresponding author}       
        
\address[unina]{organization={Department of Electrical Engineering and Information Technology,
            University Federico II},
            addressline={via Claudio 21},
            city={Naples},
            postcode={I-80125},
            country={Italy}}

\address[cnit]{organization={National Inter-University Consortium for Telecommunications (CNIT)},
            addressline={Viale G.P. Usberti, n. 181/A},
            city={Parma},
            postcode={I-43124},
            country={Italy}}





\begin{abstract}
Visible light backscatter (VLB) is an innovative
optical transmission paradigm
to enable
ultra low-power
passive communication and localization
for the Internet of Things (IoT),
by overcoming some of the limitations
of conventional (i.e., active) visible light
communication (VLC) as well as
active/passive radio-frequency (RF) technologies.
In this paper, we provide a comprehensive
survey of recent research activities in the VLB field.
After describing the principles of operation
and the main enabling technologies,
we classify the existing VLB techniques
according to several features, discussing
their merits and limitations.
Moreover, we introduce the potential
applications of VLB techniques
in several IoT domains.
Finally, we present the
main open challenges in this area and
delineate a number of future research directions.
\end{abstract}

\begin{keyword}
Backscatter \sep battery-free devices \sep
Internet of Things \sep 
metasurface, passive communication \sep
retroreflector \sep visible light communication.
\end{keyword}

\end{frontmatter}


\section{Introduction}
\label{sec:main}


The
adoption of radio-frequency (RF)
communication technologies
to support large-scale
connectivity for the Internet of Things (IoT)
exhibits severe scalability issues,
related to energy constraints,
as well as spectral efficiency limitations
and RF spectrum congestion
(so called ``spectrum crunch'').
Another drawback
of RF communications is the
leakage through walls and obstacles,
which not only complicates
interference management, but also poses
serious security and privacy concerns.
Moreover, in some scenarios,
such as aircraft/spaceship cabins and
hospitals, of hazardous environments such as
chemical or nuclear plants and oil ducts,
usage of RF technologies
must be limited or completely avoided.

To cope with the
aforementioned drawbacks,
a viable solution is to employ
\textit{visible light communication} (VLC)
techniques \cite{Pathak2015},
which work in the
portion of the wavelength
spectrum that
is visible to human eyes
(from about $400$ nm to $700$ nm),
offering a huge unlicensed bandwidth
for the potential support of a massive number of IoT
nodes.
The key idea
of VLC is to jointly carry
illumination and data,
by modulating the light emitted by
\textit{light emitting diodes} (LEDs),
which allows one to reuse
the existing lighting infrastructure
for communication or localization.

In a typical VLC deployment,
a LED, connected to the data network infrastructure via a
wired/wireless link,
performs downlink (DL)
transmission to one or multiple
client devices.
Several types of modulations,
starting from the base
\textit{on-off keying} (OOK),
can be superimposed on lighting,
provided that the switching frequency
is large enough to avoid an annoying
\textit{flickering} effect.
Visible light can also be used for
localization purposes,
leading thus to
visible light positioning (VLP)
systems \cite{Luo2017,Keskin2018,Alam2021,Majeed2021,Ma2018,Liu2022,Liu2019a}.

VLC techniques
offer several advantages
over RF ones in IoT applications.
However, a fundamental limitation of
existing VLC techniques is their inherent
\textit{one-directional}
DL transmission
(from the LED to the devices),
usually lacking a return channel
for uplink (UL) \cite{Karunatilaka2015}.
A common workaround is to employ
RF technologies in UL,
which however requires availability
of RF spectrum and must cope with
the aforementioned
scalability issues.
A more interesting solution
would be to adopt
optical communications also
for UL transmission,
such as, e.g., VLC itself, infrared (IR) or
near-ultraviolet (near-UV),
which however requires the
integration of a power-hungry
active light source (a LED or a laser)
in the client devices.
To overcome such limitations of \textit{active} VLC,
many recent papers
have resorted to
\textit{visible light backscattering} (VLB)
for ultra low-power \textit{passive}
communication and localization.

The backscatter (BS) paradigm
has been first proposed
for RF communications,
allowing a device to transmit data by reflecting
towards the source, after modulating it,
a portion of the received
electromagnetic field.
Systems based on RF backscattering
have been designed for the purposes of communication,
identification, or localization,
the most popular one being
\textit{radio-frequency identification} (RFID)
\cite{Want2006}.
A recent evolution of RF BS is
\textit{ambient backscatter}
\cite{Liu2013,Parks2014,Darsena2017,Darsena2019}
wherein passive communication
leverages existing RF signals
(such as, e.g., cellular, TV, or WiFi ones)
without requiring dedicated illuminators.


%
In this paper,
we provide a comprehensive survey of current
research activities on VLB systems, which
leverage backscatter principles
in the optical domain
to perform data communication and/or localization.
Survey papers regarding conventional
(i.e., \textit{active})
VLC are available in the literature
(see, e.g., \cite{Pathak2015,Karunatilaka2015,Rehman2019,Matheus2019}
for communication and \cite{Luo2017,Keskin2018,Alam2021,Majeed2021,Ma2018,Liu2022,Liu2019a}
for localization).
However, to the best of our knowledge,
this is the first survey paper
specifically addressing
\textit{passive} VLB systems.
In particular,
this paper has
several objectives:
\begin{enumerate}
\item
the VLB principles are introduced
and the most interesting
enabling technologies are discussed
in a simple physics-based  manner;
\item
the existing VLB techniques are presented
and their applications
in several IoT domains
are discussed,
by focusing  on weaknesses and strengths,
and the main open challenges are introduced;
\item
a number of original and promising research
developments in this field are presented.
\end{enumerate}

The paper is organized as follows.
In Section~\ref{sec:background},
a short introduction to
VLB principles is provided, and some
enabling technologies are presented in
Section~\ref{sec:tech}.
In Section~\ref{sec:channel-modeling} channel models are reviewed.
In Section~\ref{sec:state-of-the-art},
the state-of-the-art of VLB
research is discussed and the VLB
techniques are classified and
compared.
In Section~\ref{sec:applications},
potential applications of VLB techniques
to different IoT domains
are presented.
Open challenges
and future research developments are
introduced in Section~\ref{sec:future}.
Finally, conclusions are drawn in
Section~\ref{sec:concl}.


\begin{figure}[!t]
    \centering
    \includegraphics[width=0.65\columnwidth]{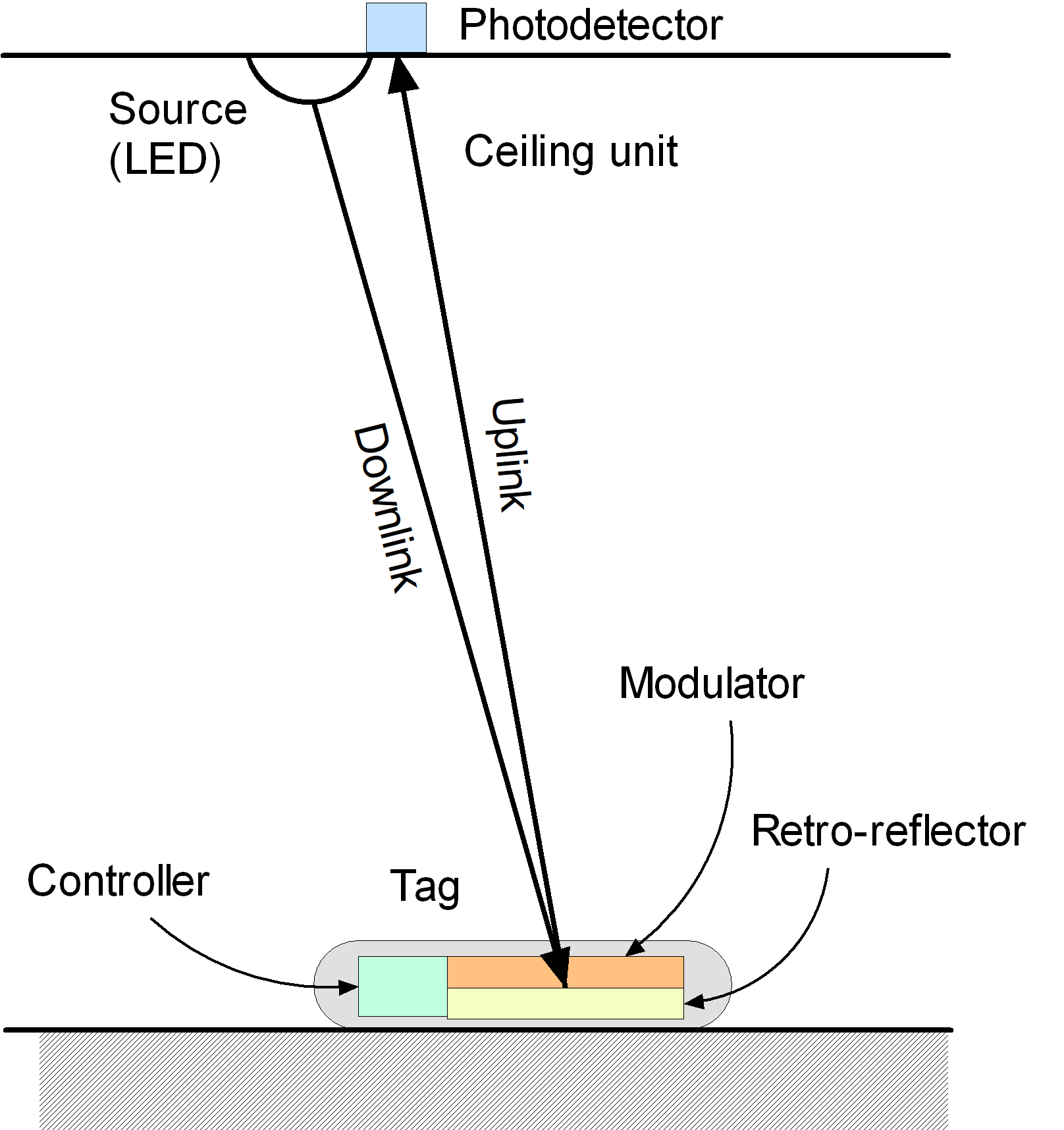}
    \caption{A pictorial view of a point-to-point VLBC system.}
    \label{fig:VLBC-system}
\end{figure}

\section{Visible light backscattering}
\label{sec:background}

We classify VLB-based systems
as \textit{visible light backscatter communication}
(VLBC) ones, targeted at
data communications, and
\textit{visible light backscatter positioning}
(VLBP) ones, aimed at localization purposes.

In a VLBC system, light is
typically used for both DL and UL
communications.
In the simplest point-to-point
scenario of Fig.~\ref{fig:VLBC-system},
a LED acts both as light source
and active DL transmitter (TX), while a device
(called a ``tag''), equipped with a \textit{retroreflector}
(see Section~\ref{sec:retroreflector}),
performs passive UL transmission
by reflecting the light,
after modulating it,
back towards the source.
Optical modulation
is  usually performed
by means of simple
\textit{LCD shutter}
devices (see  Sec.~\ref{sec:LCD-shutter}).
Due to power and complexity
constraints, very simple
\textit{intensity} modulations,
such as OOK,
are generally employed in UL, whereas
DL transmission could employ
more sophisticated spectrally-efficient
modulations \cite{Aljaberi2019}.
Cheap photodiodes (PDs)
are used at the receiver (RX)
for signal detection both in DL and UL,
even though more expensive \textit{imaging} sensors,
such as cameras,
can be employed.

VLBP systems work on similar principles,
where tag localization is usually
performed by
the transmitting LEDs
on the basis of some measured parameters,
such as the \textit{time-of-arrival} (TOA),
\textit{time-difference-of-arrival} (TDOA)
or \textit{received signal strength} (RSS),
together with some information backscattered
by the tag itself, such as the tag identity,
timestamps, or the transmitted power.
Such VLBP systems are generally used indoors,
where the GPS signal is often not available.
One distinct advantage
of VLBP  techniques over active/passive RF ones,
like those based on WiFi,
is that the number of LED luminaries and their power
is generally much higher that that
of WiFi access points \cite{Pathak2015}.

VLB propagation links can be classified according to the degree of directionality of
source and tag, and  the existence of a {\em line-of-sight (LOS) path} between them \cite{Wang2017}.
{\em Directed links} employ sources and tags with narrow
radiation and illumination patterns, respectively,
which have to be aligned in order to minimize
path loss effects and reception of ambient light noise.
On the other hand, {\em nondirected links} employ wide-angle
sources and tags, which do not require such an accurate alignment.
{\em Hybrid links} are also possible, which combine sources and
tags having different degrees of directionality.

The presence of a LOS path between the source and the tag
allows link designs with maximum power efficiency and
minimum multipath distortion.
{\em Non-line-of-sight (NLOS) links} - also referred to as
{\em diffuse links} - generally
rely upon reflection of the light from the ceiling or some
other diffusely reflecting surface, such as
people or cubicle partitions, stands between the source
and the tag. Compared to their LOS counterparts, NLOS link designs ensure
greatest robustness and ease of use
(see also Section~\ref{sec:channel-modeling} for
further details).

\section{VLB-enabling technologies}
\label{sec:tech}

In this section we describe
two main enabling technologies
for VLB system:
\textit{retroreflectors}
and \textit{LCD shutters}.


\subsection{Retroreflectors}
\label{sec:retroreflector}

An optical retroreflector (RR)
is a device that,
unlike a mirror,
reflects the incident light
back towards the direction
of the source,
with minimal scattering.
RRs can be implemented with different
technologies and are used in
many fields,
including
free-space optical
communications networks
\cite{Zhou2003,Junique2006,CarrascoCasado2011},
satellite communications \cite{Le2018},
and low-powered sensor networks \cite{Hsu1998,Teramoto2004,Khalid2018}.
Cheap RR materials are commonly available
(e.g., Scotchlite\textsuperscript{\texttrademark}
manufactured by 3M)
and are used for road signs, bicycles, and clothing
for road safety \cite{Xu2017}.

%
\begin{figure}[!t]
    \centering
    \includegraphics[width=0.75\columnwidth]{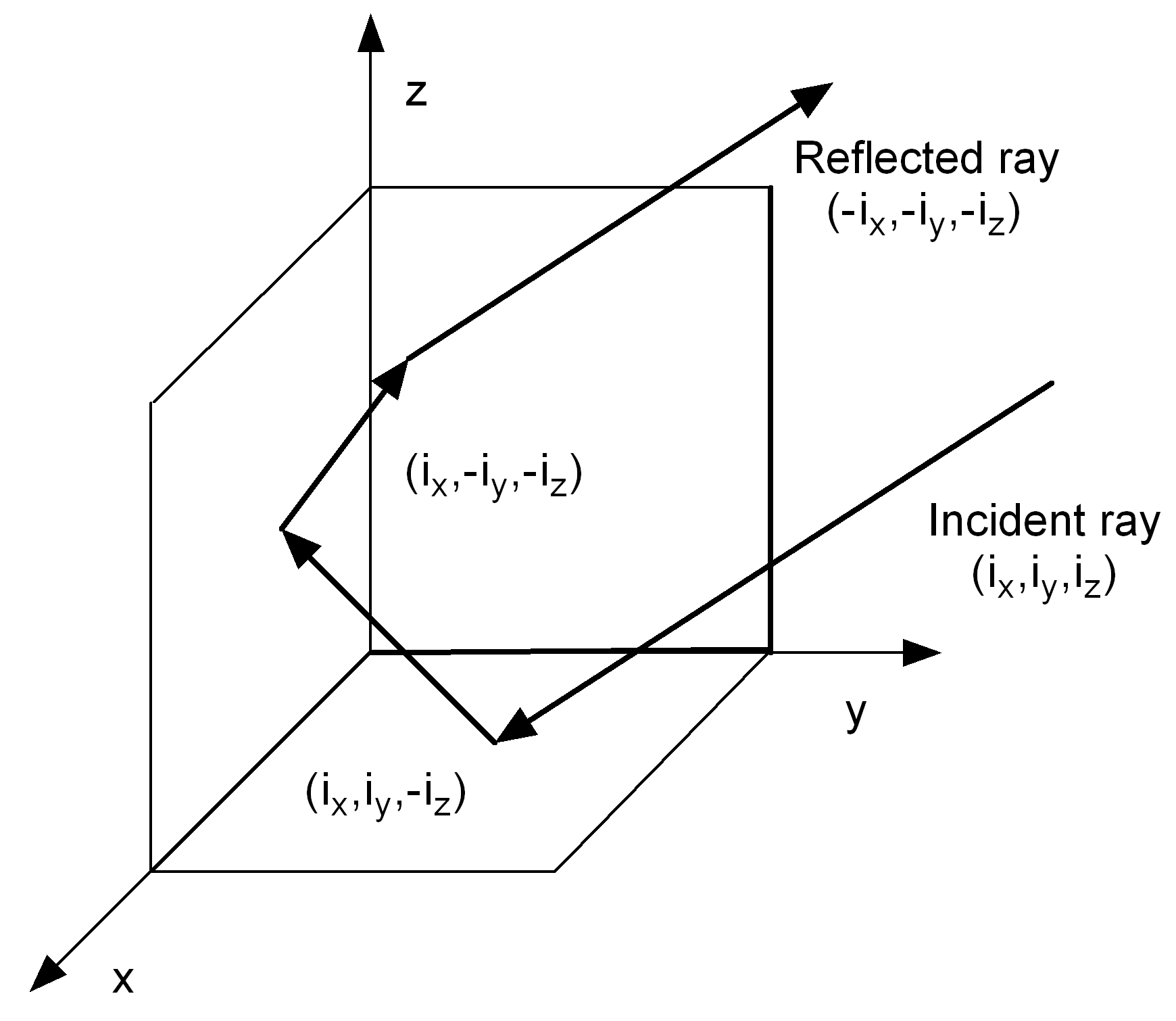}
    \caption{Working principle of a CCRR based on ray optics.}
    \label{fig:RR-3D-cornercube}
\end{figure}
%

One popular type of RR is the
\textit{corner-cube retroreflector} (CCRR)
\cite{Janik2017},
which is composed by three mirrors
arranged in a $90\degree$ corner geometry.
Light rays are sent back towards
the source regardless
of the relative orientation of the
incoming beam direction, after undergoing
three reflections,
as depicted in Fig.~\ref{fig:RR-3D-cornercube}.
Another common type of RR
is the \textit{spherical retroreflector}
(commony known as ``cat's eyes''),
which is build as a
high index-of-refraction
transparent sphere with a reflective
backing \cite{Arecchi2007}.
A less common implementation is is the \textit{phase-conjugate mirror}
\cite{Eichler2012}, which
exploits nonlinear optics phenomena such as
four-wave mixing or stimulated Brillouin scattering.
It is worth noting
that innovative materials
such as \textit{metasurfaces}
can also be used as RRs,
as further discussed in
Section~\ref{sec:metasurfaces}.

\begin{figure}[!t]
    \centering
    \includegraphics[width=0.85\columnwidth]{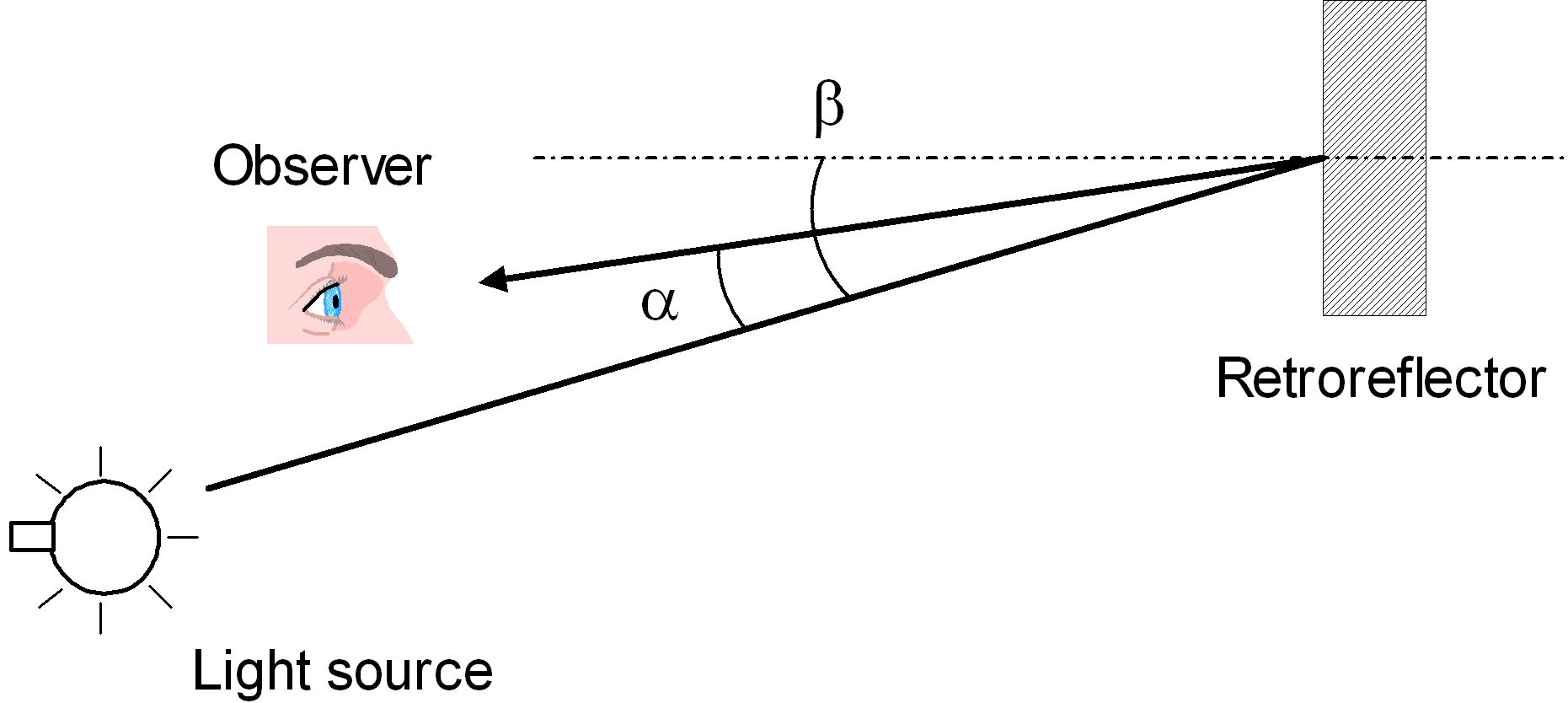}
    \caption{Geometrical description of a RR: $\beta$ is the entrance angle, $\alpha$ is the
    observation angle.}
    \label{fig:RR-geometry}
\end{figure}

Disregarding the actual implementation,
it is common to describe
\cite{Arecchi2007}
the behavior of a RR device
by means of two angles (Fig.~\ref{fig:RR-geometry}):
the \textit{entrance angle} $\beta$,
which is the angle between the illumination direction
and the normal to the RR surface,
and the \textit{observation angle} $\alpha$,
which is the angle between the illumination direction
and the viewing direction.
High-quality retroreflectors
work over fairly wide entrance angles, up to $45\degree$
or more (up to $90\degree$ for pavement marking),
with very small observation angles
($< 1\degree$).

The performance of RRs can be
measured by several coefficients,
the most common ones \cite{Arecchi2007}
are $R_I$ and $R_A$.
The first one is the
\textit{coefficient of retroreflected luminous intensity:}
\be
R_I = \frac{I}{E_{\perp}} \qquad \text{[cd/lux]}
\ee
where $E_{\perp}$ is the illuminance (in lux)
on a plane normal to the direction of illumination,
and $I$ is the intensity (in cd)
of the illuminating light.
The second one is the
\textit{coefficient of retroreflection}:
\be
R_A = \frac{I}{E_{\perp} A} = \frac{R_I}{A}
\qquad
\text{[(cd/m$^2$)/lux]}
\ee
where $A$ is the area of the retroreflector.
Values for RA of several hundred (cd/m$^2$)/lux are not
uncommon \cite{Arecchi2007}.
Both coefficients are functions of the
angles $\beta$ and $\alpha$.

It should be noted that RRs can also be
used as optical modulators,
by controlling the reflection
mechanisms
with \textit{micro-electromechanical systems} (MEMS)
\cite{CarrascoCasado2011}
or semiconductor \textit{multiple quantum wells} (MQW)
technologies \cite{Xu2017}.
%


\subsection{LCD shutters}
\label{sec:LCD-shutter}

LCD shutters are
employed in consumer 3D TV glasses
\cite{Yun2016} and
act as modulator devices in
most VLB protoypes.
%
An LCD shutter is characterized by a multilayer
sandwich structure \cite{Li2015},
with two linear polarizers
at the ends and one
\textit{liquid crystal} (LC)
layer in between.%
\footnote{A liquid crystal (LC) is a special material
whose properties are between those
of a liquid and those
of a crystal. The most common type of LC is a \textit{nematic}
liquid crystal, whose optical behavior can be modified by applying
an electric field to it.}
Each polarizer obeys the \textit{Malus law:}
\be
I_{\theta} = I_{0} \, \cos^{2} \theta
\ee
where $I_{0}$ is the intensity
of light impinging on the polarizer,
$I_{\theta}$ is the intensity
of light that is passed
through the polarizer, and
$\theta$ is the angle
between the polarization direction
of the polarizer and the polarization
of light.
If the polarization of the incident
light is parallel to that of the
polarizer, i.e., $\theta =0\degree$,
no attenuation will occur and the light
will pass unaltered through the polarizer.
When, instead, the two
directions are perpendicular, i.e.,
$\theta = 90\degree$, the entire incident
light is blocked by the polarizer.

\begin{figure}[tb]
    \centering
    \includegraphics[width=0.75\columnwidth]{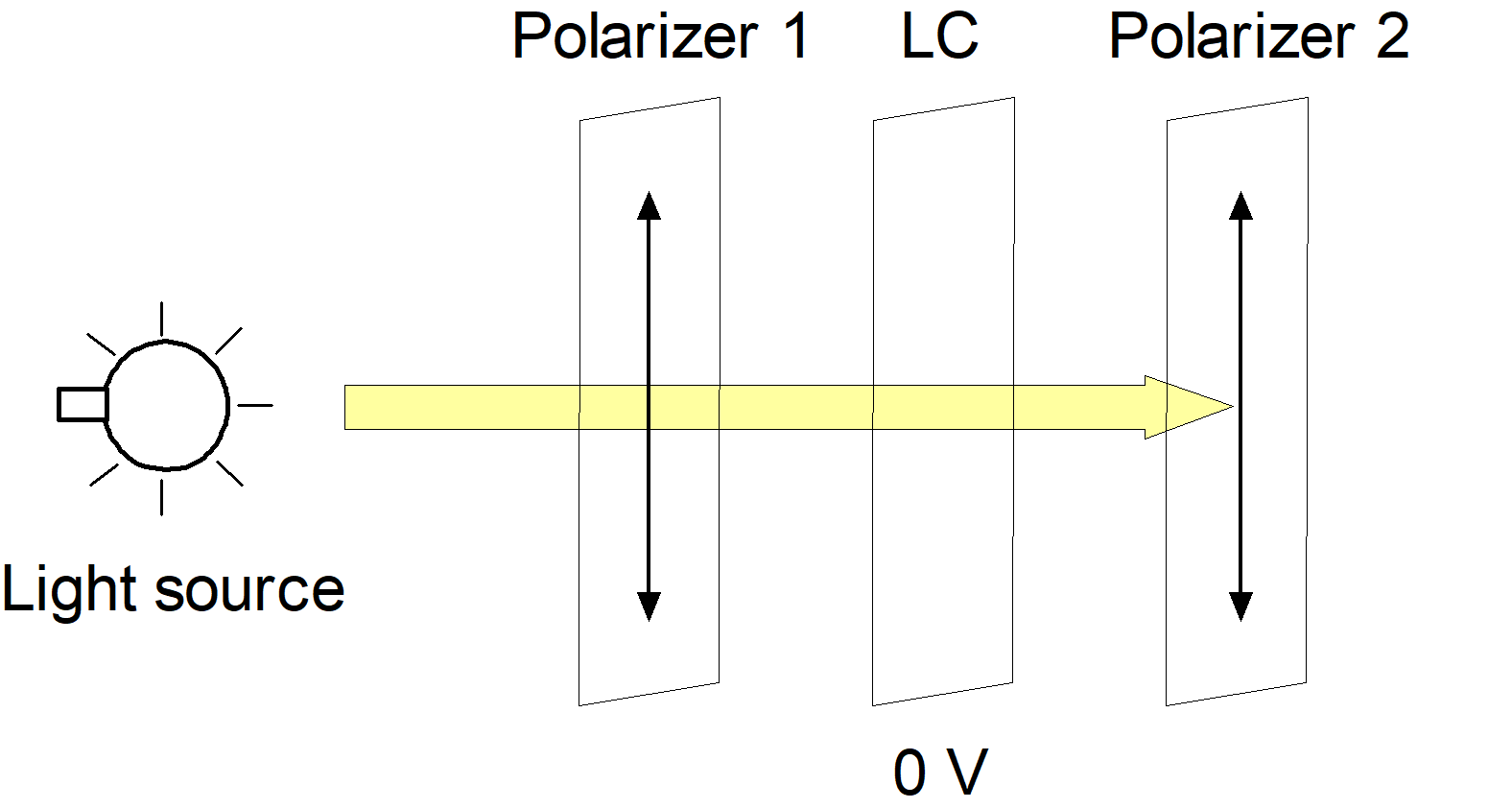}\\[10mm]
    \includegraphics[width=0.75\columnwidth]{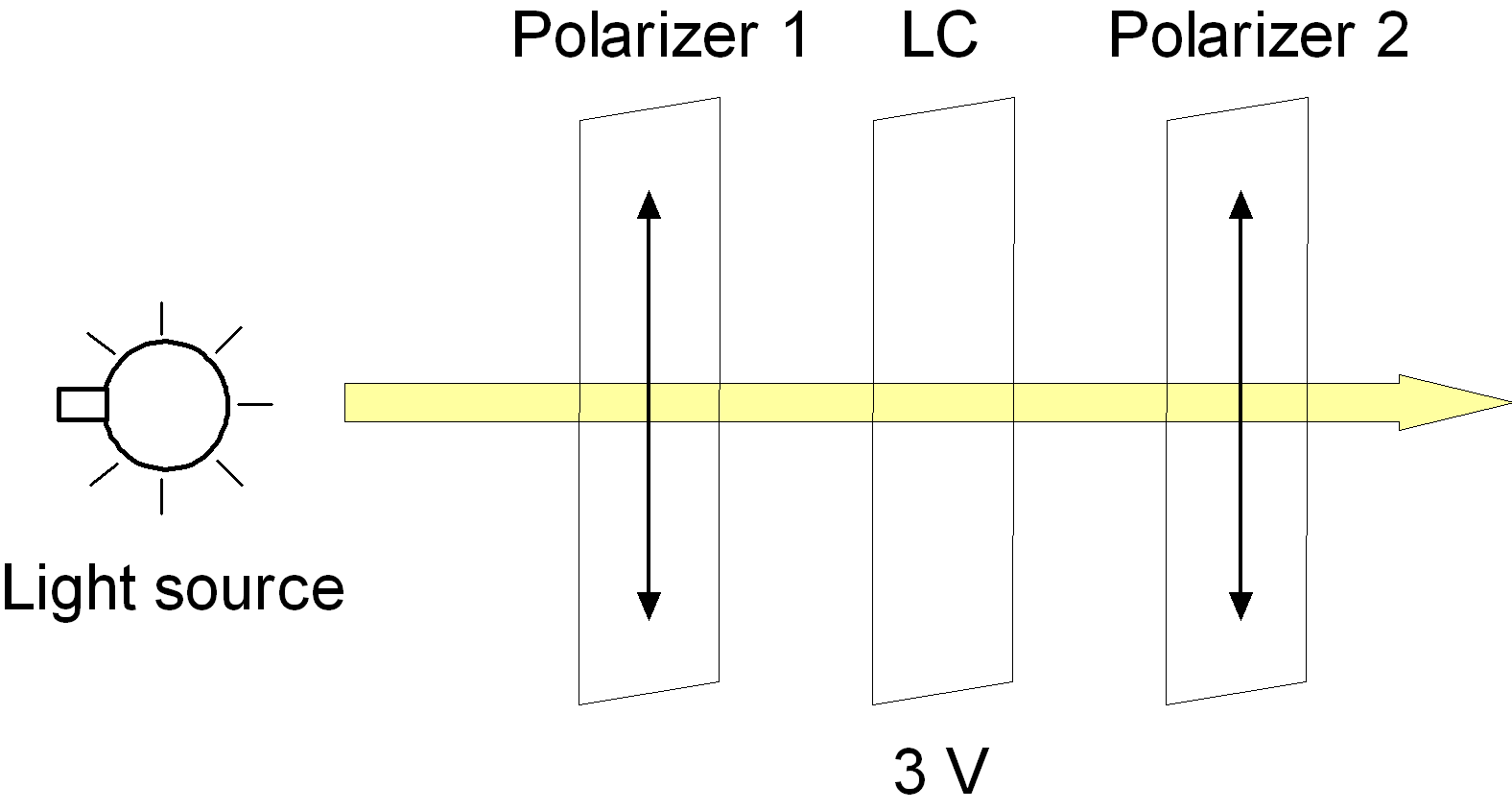}
    \caption{An LCD shutter operating as an OOK modulator.
    When no voltage is applied to the LC
    (top), the polarization of the light is rotated by 90 degrees
    and is blocked by the second polarizer.
    When a certain voltage ($3$ V) is applied to the LC (bottom),
    the polarization of the light is not rotated and passes trough the second
    polarizer.}
    \label{fig:LCD-shutter}
\end{figure}

Modulation of bits in a LCD shutter
is based on polarization
properties of light.
In Fig.~\ref{fig:LCD-shutter},
the working principle of
a \textit{twisted nematic} (TN)
LCD shutter is explained \cite{Yang2019}.
A variable voltage is applied between
the layers of a TN LC,
which determines
a twist/untwist (realignment)
of the LC molecules.
In most commodity TN LCs,
when no voltage is applied (normal or uncharged state),
a light beam passing through the LC layer
undergoes a rotation by $90\degree$.
Thus, polarizer 2 can entirely
block the light if placed
parallel to the incident light.
Therefore, when the voltage is increased,
untwisting of LC molecules
will happen and polarizer 2
will become brighter.
Over a certain voltage level,
such as $3$ V,
complete untwisting
of LC molecules
occurs (charged state)
and this result in the brightest
state of polarizer 2.
Hence, the intensity of the light can be modulated
by tuning the applied voltage
so as to encode bits using
bright and dark states.%
\footnote{Some LCD shutters employ orthogonal
polarizers at the two ends \cite{Wu2020},
in which case the encoding of voltage
levels to bright/dark states
is simply reversed.}

The most interesting
features of LCD shutters are, among others,
compatibility with non-coherent white light \cite{Wang2019},
low power consumption (sub-mW \cite{Xu2017})
and low cost ($0.03 \, \$$ per cm$^2$ \cite{Yang2015}).
These features make LCD shutter the best choice
for large scale IoT deployment,
in comparison to RR-based modulators.
However, LCD shutters exhibit
a slow (in the order of ms) and asymmetric impulse response
\cite{Yun2016,Wu2020},
which limits the achievable data-rate
to sub-kbps level and complicates equalization.
In \cite{Yun2016} a rising time of $1.096$ ms and a
falling time of $0.533$ ms have been measured,
whereas in \cite{Wu2020} the corresponding values
of $0.3$ ms and $4$ ms were found.
The rising time is much smaller than the falling time
because of the slow discharging property.

A simplified model for the response
of the LCD shutter is that of a first-order
RC filter with a cutoff frequency
in the order of $300$ Hz \cite{Yun2016}.
However, LCD shutters exhibit an
asymmetric response
and a marked nonlinear behavior \cite{Kim2015,Wu2020},
which make them very difficult
to generate multilevel modulated signals
\cite{Xu2017}.
Finally, the slow switching rate of LCD shutters
might induce flickering.


\section{Channel modeling for wireless optical communications}
\label{sec:channel-modeling}

{
A considerable amount of work has been carried out
about channel characterization, covering both experimental measurements and computer
modeling of indoor and outdoor optical wireless communications systems.
In this section, we focus  on indoor scenarios, which are in our opinion the most appealing ones
for VLB techniques.
For the characterization of outdoor optical wireless channels we directly refer to \cite{Ghassemlooy2018}.}

{The channel characteristic of an optical wireless link is fixed for
a given position of the source, tag, and surrounding reflecting objects.
It changes only when these components are moved by distances
in the order of centimeters \cite{Kahn1997}. Due to high
bit rates and the relatively slow movement of objects and people
within a room, the optical wireless channel typically varies
only on a time scale of many bit periods and, hence, it may be considered quasi-static \cite{Street1997}.}

{For LOS links, reflections are negligible and, consequently, the
path loss is easily calculated from the knowledge of the transmitter beam divergence, receiver size,
and separation distance between the source and the tag \cite{Kahn1997}.
NLOS links, particularly in indoor applications, are subject to the effects of
multipath propagation, similarly to RF systems.
Multipath propagation causes the received signal to suffer from severe amplitude fades on
the scale of a wavelength. However, optical receivers are typically equipped with
large-area PDs, whose surface area is
orders of magnitude larger than the transmission wavelength.
In this case, the total photocurrent generated at the receiver
is proportional to the integral of the optical power over such a large surface,
thus providing an inherent spatial diversity
that averages out fading effects \cite{Kahn1997}.
Although indoor optical links are inherently robust against the effects of multipath fading,
they suffer from the effects of multipath-induced dispersion, which causes
{\em intersymbol interference (ISI)}, thus adversely affecting link performance.}

{Several experimental channel characterizations have been
carried out \cite{Zwillinger1988,Xiang2014,Dong2017,Chvojka2017,Jovkova2000,
Carruther2000}, by considering both LOS and NLOS configurations.
Results have shown that the optical channel response is sensitive not only to the
position of the PD, but also to its orientation and rotation.
Different models have been proposed, e.g., ceiling bounce \cite{Kahn1995},
Hayasaka-Ito \cite{Hayasaka2007}, and spherical \cite{Jungnickel2002}, which
are however targeted at the
{\em infrared (IR) region} of the electromagnetic spectrum.
An accurate channel characterization for VLB applications
is still lacking (see Section~\ref{sec:future}).}

{Optical transceivers operating in typical indoor environments are subject to intense {\em ambient light},
emanating from both natural and artificial sources. The main sources of ambient light are the
sunlight, incandescent lamps, and fluorescent lamps.
The DC background photocurrent generated by the ambient light acts as a noise
source in the receiver, referred to as {\em shot noise}, which degrades the link performance.
The sunlight is typically the strongest source of shot noise and
represents an unmodulated source of the ambient light with a very wide spectral width and a
maximum power spectral density (PSD) located at about $500$ nm.}

{All artificial ambient light sources are modulated, either by the mains frequency or, in the
case of some fluorescent lamps, by a high-frequency switching signal.
The background current due
to artificial illumination is only a few tens of $\mu$A, which is well below that produced by sunlight, which
could be as high as $5$ mA \cite{Moreira1995}.
Incandescent lamps have
a maximum PSD around $1$ $\mu$m and produce an interference signal, which is a near-perfect sinusoid with the same frequency of the AC grid. Only the first few harmonics carry a significant amount of energy, and interference effects can be effectively reduced by using a high-pass filter (HPF)
\cite{Moreira1997,Moreira1996a,Moreira1996b}.
The interference produced by fluorescent lamps driven by conventional ballasts is a
distorted sinusoidal and extends up to $20$ kHz. The spectrum of the interference depends on the
switching frequency. In this case, the interference can be modeled using a high-frequency component and a low-frequency component, and it can be effectively reduced
using an HPF with a cut-off low frequency \cite{Moreira1996a}.}

{Other noise sources like IR audio headphone transmitters and a TV remote control unit can be
viewed as sources of the noise as their operating wavelengths of coincide with optical
wireless transceivers \cite{Hayasaka2007}.
The aforementioned studies again pertain IR communications
and the impact of ambient light on VLB systems deserves further investigation
(see Section~\ref{sec:future}).
Another form of noise - which is distinctive of VLB systems - is represented by
the \textit{auto-interference}: the illuminating
source acts as a strong interference on the weak
backscattered signal
at the UL RX.
To mitigate such a phenomenon,
some form of shielding
can be used \cite{Li2015}. Time-varying
metasurfaces can also be an innovative solution
for countenancing auto-interference (see Section~\ref{sec:metasurfaces}).}


\section{Existing VLB techniques}
\label{sec:state-of-the-art}

In this section, a survey of the existing
VLB techniques is provided, whose
main features are also summarized
in Tab~\ref{tab:tab1}.

\smallskip
\subsubsection{Retro-VLC \cite{Li2015}}

This bi-directional VLBC system
performs DL transmission
by employing
conventional VLC techniques, whereas
UL transmission is based on VLBC.
The UL TX employs a
RR device coupled with
an {LCD shutter} as modulator.
Several power optimization solutions
are proposed, together with
an implementation
based on purely analog techniques
to reduce energy consumption at the tag.
The Retro-VLC prototype employs
Manchester-encoded
binary signals, achieving a data-rate
of $10$ kbps in DL and $0.5$ kbps
in UL over a range of $2.4$ m in
an office environment.

\smallskip
\subsubsection{Ambient light BS communication \cite{Yun2016}}

This VLBC system  exploits either ambient light
or a dedicated illuminator for UL transmission.
The design is similar to
\cite{Li2015} and the developed
prototype achieves
a data-rate of $100$ bps
over a $10$ cm range.
Another VLBC system for UL communications
is devised in \cite{Wang2016},
where information is statically
encoded in a tag composed
by a reflective surface.
Such a surface is
mounted on a mobile object (like a car or truck),
and can be read by illuminating it with an
unmodulated source (artificial or natural)
and detecting the reflected light
with a PD-based RX.
The system works similarly
to a bar-code reader but without
using energy-consuming
cameras for reading,
since the power
consumed by the PD
is very low (about $1.5$ mW).
One possible improvement
of the system proposed in \cite{Wang2016}
is the dynamic encoding of data,
which can be performed by adopting
advanced materials (such as e-ink screens
or LCD shutters)
whose reflection properties
are adjusted in real-time.

\smallskip
\subsubsection{Pixelated VLC-backscatter \cite{Shao2017}}

The VLBC scheme proposed
in \cite{Shao2017}
employs multilevel modulation
to improve the data-rate of the simple OOK-based
Retro-VLC technique \cite{Li2015}.
Since the highly nonlinear
characteristic of the LCD shutter
prevents modulating a single LCD-shutter
with more than two levels,
the proposed solution
employs \textit{multiple} LCD shutters,
whose number is equal to $\log_2(M)$
(with $M$ denoting the cardinality)
of the modulation, which are
independently switched by
the modulation bits.
A RR and a LCD
shutter form a \textit{pixel}
that can be switched
independently from the others.
Hence, the overall
reflected light is
proportional to
the number of activated pixels,
which allows one to obtain multilevel
pulse amplitude modulation (PAM)
signals.
The prototype in \cite{Shao2017}
employs up to three pixels and works at the
symbol rate of $200$ symbols/s,
implementing OOK, 4-PAM, and 8-PAM modulations,
with an achieved throughput
of $600$ b/s at $2$ m,
$400$ b/s at $3$ m,
and $200$ b/s at $5$ m.
Rate adaptation should be performed,
where lower-order and more robust
modulations, such as OOK,
are used when the range is higher
and viceversa.
A possible improvement to this scheme,
mentioned by the same authors,
is the adoption of orthogonal frequency-division multiplexing (OFDM)
modulation techniques.

\smallskip
\subsubsection{PassiveVLC \cite{Xu2017}}

The technique proposed in \cite{Xu2017}
tries to improve the design of Retro-VLC \cite{Li2015}
by specifically
focusing on the modulation/coding schemes.
In particular, to solve the problem
of long consecutive
stream of zeros/ones,
it is proposed to replace the meoryless
Manchester coding scheme employed in \cite{Li2015}
with the Miller one, due to its better spectral efficiency.
The choice of a modulation scheme
with memory slightly
complicates the decoding algorithm,
which can be formulated
as an optimization problem and solved
by means of dynamic programming methods.
Another innovation
of the PassiveVLC technique
is avoiding to completely
charge/discharge the LCD shutter, in order to
reduce the switching time
(from $4$ ms down to $1$ ms).
The resulting solution is called
\textit{trend-based modulation} (TBM),
since the information
is mapped on the
``trend'' of the voltage change,
avoiding hence the slower complete
charge/discharge,
at the price of
a reduced immunity to noise.
The PassiveVLC prototype
achieves up to $1$ kbps in UL over $1$ m
for a flexible range of orientations
and different ambient light conditions.

\smallskip
\subsubsection{RetroArray \cite{Wu2019}}

This technique exploits
the nonlinear behavior
of the LCD shutter to improve the data-rate
of existing VLBC systems.
Indeed, LCD shutters exhibit
an highly asymmetric response,
with $0.5$ and $4$ ms charging and discharging time,
respectively.
The RetroArray prototype employs
an array of LCD shutters
and encodes the information only on the
rising (i.e., charging) edges,
resorting to time interleaving between
different shutters
to achieve multilevel modulation.
The technique, called
\textit{delayed superimposition modulation} (DSM),
allows one to achieve an UL data-rate
of $4$ kbps over $3$ m
using an array of $16$ LCD shutters.

\smallskip
\subsubsection{Poster \cite{Wang2019}}

It employs
\textit{polarization-based quadrature-amplitude modulation}
(PQAM) to overcome the problem
of flickering, associated to the
low switching rate of
the LCD shutter.
Similarly to \cite{Yang2015},
the LCD TX employs
only one polarizer, thus
mapping the information
on two orthogonal polarizations,
which are controlled
by the LC layer.
This completely avoids
flickering, since
the intensity of
the backscattered light
is not changed.
At the RX, a second polarizer
is capable of detecting the
different polarizations.
The PQAM design is robust to the cases
where TX and RX
are not perfectly aligned, in terms
of polarization angle.

\smallskip
\subsubsection{RetroI2V \cite{Wang2020a}}

This system exploits retroreflective
coating of the road signs for
VLBC in outdoors scenarios.
Late-polarization and polarization-based differential
reception technique
are used to mitigate flickering.
Experimental results show that the system exhibits
long range (up to $100$ m) connectivity.
Efficient operation is achieved by using
a decentralized multiple-access control (MAC)
protocol.
A limitation of this scheme
is that it can be used only for one-way communication
and it is designed
for a specific vehicular
application.

\smallskip
\subsubsection{RetroTurbo \cite{Wu2020}}

A VLBC prototype
is implemented in \cite{Wu2020},
which uses PQAM and DSM for IoT-oriented applications.
It achieves an UL data-rate of $8$ kbps at $7.5$ m range,
which is $32$x higher than OOK \cite{Li2015}.
The range increases to $10$ m if the data-rate
is decreased to $4$ kbps.
The Authors also designed  a real-time
demodulation algorithm
and claimed a $128$x rate gain ($32$ kbps)
via emulation.

\smallskip
\subsubsection{RETRO and PassiveRETRO \cite{Shao2018,Shao2019,Shao2020}}

In \cite{Shao2018,Shao2019,Shao2020}
several VLBP systems for localization
are proposed.
The RETRO system \cite{Shao2018,Shao2020}
performs real-time tracking of location
and orientation of passive IoT nodes equipped with
a RR and an LCD shutter, which
transmit their identity to the network.
The prototype exploits received signal strength (RSS)
measures and trilateration to
achieve ultra low-power
centimeter-level positioning.
An improvement is the RR-based
PassiveRETRO system \cite{Shao2019},
which retains the advantages
of RETRO \cite{Shao2018,Shao2020} but
eliminates the LCD shutter at the tag.
This design choice completely avoids
the necessity of any electronic component
on the IoT devices.
Polarization-based modulation and
bandpass optical filters
are used as means to identify
and separate the signals reflected by different
IoT devices.
Moreover, \textit{optical rotatory dispersion}%
\footnote{Optical rotatory dispersion is a property of some materials
that rotate the polarization of light
to different extents depending on the wavelength.}
is used for mitigating
interchannel interference and improving
signal-to-interference-plus-noise ratio (SINR)
for each node.
Specifically, the PassiveRETRO system
splits the LCD shutter present in RETRO
into two parts: one linear polarizer and
one LC layer are installed
on the light source for performing
polarization-based modulation,
while another linear
polarizer is placed on top of the
RR to modulate
the backscattered light.


\begin{sidewaystable*}[!th]
{
\scriptsize
\caption{A summary of existing VLB techniques}
\begin{tabular}{p{20mm}p{18mm}p{18mm}p{15mm}p{10mm}p{15mm}p{18mm}p{32mm}p{28mm}}
\toprule
{Scheme} & {Technology} & {Modulation} & {UL data-rate}       & {Range}
& {Tag power}    & {Applications} & {Main advantages}  & {Main limitations} \\
         &              &              & \hspace{3mm} [bps]   &  \hspace{1.3mm}[m]
& {consumption}  &                &                 &           \\ \midrule
Retro-VLC \cite{Li2015}
& RR + LCD shutter
& OOK w/ Manchester coding
& $500$
& $2.4$
& $234$ $\mu$W
& Communication (IoT)
& Bidirectional, low-power, analog implementation
& Low-rate, narrow field-of-view
\\ \midrule
Ambient light BS communication \cite{Yun2016}
& RR + LCD shutter
& OOK
& $100$
& $0.1$
& N/A
& Communication (IoT)
& Ambient light source or dedicated illuminator
& Low-rate, low-range
\\ \midrule
Ambient light passive
communication \cite{Wang2016}
& Reflective surface
& OOK w/ Manchester coding
& $50$  & $1$ & N/A
& Identification and tracking (IoT)
& Ambient light source, mobile node
& Very low-rate, low-range, static encoding of information
\\ \midrule
Pixelated VLBC \cite{Shao2017}
& Multiple LCD shutters
& OOK, 4-PAM, 8-PAM
& $600$
& $2$
& $200$ $\mu$W
& Communication (IoT)
& Multilevel modulation, high data-rate, rate adaptation
& Complexity
\\ \midrule
PassiveVLC \cite{Xu2017}
& RR + LCD shutter
& TBM w/ Miller coding
& $1000$
& $1$
& $150$ $\mu$W
& Communication (IoT)
& High data-rate,
flexible range of orientations
& Reduced immunity to noise
\\ \midrule
RetroArray \cite{Wu2019}
& Multiple LCD shutters
& DSM
& $4000$  & $3$
& N/A
& Communication (IoT)
& High data-rate
& Complexity
\\ \midrule
Poster \cite{Wang2019}
& RR + LCD shutter
& PQAM
& N/A
& $2$m
& N/A
& Communication (IoT)
& Flicker-free, robust to TX/RX orientation
& Complexity
\\ \midrule
RetroI2V \cite{Wang2020}
& Retroreflective coating
& OOK
& N/A
& $\sim 100$
& N/A
& Communication (I2V)
& Decentralized MAC protocol, flicker-free, long range
& One-way communication, only vehicular scenario
\\ \midrule
RetroTurbo \cite{Wu2020}
& N/A
& PQAM + DSM
& $8000$
& $7.5$
& $800$ $u$W
& Communication (IoT)
& High data-rate, real-time demodulation algorithm
& Complexity
\\ \midrule
RETRO \cite{Shao2018,Shao2020}
& RR + LCD shutter
& N/A
& N/A
& $1.5$
& Ultra low-power
& Localization (IoT)
& Based on RSS and trilateration,
centimeter-level positioning
& N/A \\ \midrule
PassiveRETRO \cite{Shao2019}
& RR + linear polarizer
& PQAM
& N/A
& $1$
& Ultra low-power
& Localization (IoT)
& Mitigation of interchannel interference,
completely passive tag
& RX complexity
\\ \bottomrule \\
\multicolumn{9}{p{200mm}}{\vspace{-4mm}
DL=downlink, UL=uplink, RR=retroreflector,
DSM = delayed superimposition modulation,
TBM = trend-based modulation
}
\end{tabular}
\label{tab:tab1}
}
\end{sidewaystable*}


\section{Main applications of VLB in IoT}
\label{sec:applications}

VLB techniques can be employed
in several IoT domains,
mainly to supplement or
replace VLC and/or RF technologies
when extreme energy efficiency is pursued.
In the following, some
fields of application are discussed.


\subsection{Healthcare}

In e-Health and m-Health
applications \cite{Rodrigues2016,Yang2022},
sensors of different nature are used
to monitor some physiological
parameters of the patients
(such as, e.g., temperature, pulse, blood
pressure, or oxygen saturation)
and transmit them in real-time
to a collection unit.
The use of RF communication technologies
to this aim presents
two main drawbacks:
(i) long-term overexposure to RF fields,
which amplifies the risks to human health;
(ii) electromagnetic interference (EMI),
which affects the accuracy and reliability
of medical equipments.
Optical-based VLC and VLB techniques
allows one to overcome the previous drawbacks,
and can be used in many different
healthcare setups, including
operating and emergency rooms,
intensive care units,
imaging and pathology labs,
and hospital wards.
In particular, the use of
VLB techniques
is particularly
appealing in \textit{wireless body area networks}
(WBANs) \cite{Haddad2020},
composed by wearable or textile sensors
aimed at long-term health monitoring.
Indeed, WBAN must inherently adopt
energy-efficient devices and
protocols for sensing and communication.
Moreover, wearable devices are equipped
with low-capacity batteries,
whose recharging or substitution
might be cumbersome.

In \cite{Noonpakdee2013},
a VLBC system for health monitoring
applications is studied, which
exploits the light
emitted by a LED to transmit,
by means of a CCRR,
OOK-modulated data
acquired by wearable sensors
to a central unit.
The RX at the ceiling
employs an imaging
sensor to detect the
transmitted data.
A link budget analysis
is proposed, aimed at
assessing the theoretical performances
of the system, in terms of BER,
achievable range, and data-rate.
This solution
is further generalized in \cite{Noonpakdee2014,Noonpakdee2016}
to an \textit{hybrid} system
providing two different
operation modes:
an \textit{active} one based on IR and a \textit{passive}
one employing VLBC.
The active mode is used when
VLBC cannot be used, that is,
when the LED light is turned
off or the user is too far
from the source.


\subsection{Transportations}

{Intelligent transportation systems
(ITSs) are critical components
to make transport safer, more
efficient, more reliable,
and more sustainable.
They make widespread use
of IoT technologies
to enable automated collection of
transportation data and
information exchange
between vehicles/passengers
and infrastructures.}

{Recently,
VLC techniques have been proposed
to replace RF ones
in \textit{vehicle-to-vehicle} (V2V),
\textit{vehicle-to-infrastructure} (V2I), and
\textit{infrastructure-to-vehicle }(I2V) links,
(see \cite{Eldeeb2022} and references therein).
Their use relies on
the ubiquitous availability of
LED-based street, traffic and vehicle lights.
A distinctive feature of
VLC-based vehicular communications
is the outdoor operating scenario,
characterized by a
non-negligible
ambient light interference
due to background solar radiation.
As discussed in Section~\ref{sec:channel-modeling},
this type of interference
can adversely affect
the reliability of VLC and
VLB techniques, unless suitable
mitigation strategies are performed.}

{The VLBC system called
RetroI2V \cite{Wang2020a}
assures flicker-free I2V data transmission
over distances of about $100$ m, under
different lighting conditions.
It works by equipping the road signs with
tiles of transparent LCD shutters
plus controling/harvesting
circuitry, actually converting
conventional road signs into
smart, dynamic transmitters.
Several usage scenarios
are discussed in \cite{Wang2020a},
aimed at providing additional information
to drivers (i.e., possible
accidents or time restrictions)
or adapting the messages
to road/weather conditions.}

{Notwithstanding the successful demonstration of
\cite{Wang2020a},
we do not envision
VLB as a candidate technology
for outdoors mission-critical
applications, such as autonomous
driving or collision
avoidance systems.
VLB can  be used
instead to transmit or share
low-rate information,
such as traffic or road conditions,
points-of-interest, or travel suggestions.
in non-critical infotainment
applications.
Other areas where VLB technologies can be fruitfully
used is in automatic toll or ticketing
systems, to provide information
about free parking lots, or
as support to green shared-mobility systems, such as
bikesharing or scooter sharing systems.}

{We envisage that VLB technologies
are more useful in public transportation systems
working in indoor scenarios,
such as galleries or metro railway stations,
or to provide low-rate communications
within vehicles.
For example, VLB can be used to provide
indoor localization when GPS is absent,
or to assist precise stop control
of train vehicles to platforms.
Another application of is substitution
of RF techniques in smart ticketing and
access control systems.
Moreover, VLB technologies
can be used also to support more innovative
functions, such as counting people
to implement crowd-avoiding functionalities in ITSs \cite{Darsena2020}.}


\subsection{Smart cities}

{The wide
availability of outdoor lightning infrastructures
in urban environments
is a formidable enabler for VLC and
VLB applications.
Moreover, adoption of VLB-based sensing
and communications can avoid to further congestionate
the RF spectrum in urban
environments.
Possible application of interest could be,
for instance, in smart parking (detection of free spaces in parking lots),
environmental sensing (i.e., for pollution checking)
and cultural heritage.
To enable long-range communications, it is envisioned that
unmanned autonomous vehicles (UAVs) can be used to
relay the information gathered by VLB sensors
to a central unit \cite{Ji2019}.}


\subsection{Smart home}

{IR-based device control is common in
consumer electronics and equipment:
however, IR devices are subject
to annoying battery replacement.
Since homes and offices
are equipped with LED luminaries,
IR active devices can be easily replaced
by passive VLB system, which assure
long times of operation with a limited
energy consumption.}

{Moreover, the indoor positioning capabilities of
VLB  make it the naturale candidate for
\textit{home robotics} applications \cite{Sahin2007} (vacuum cleaners,
monitoring devices, lawn mowers, etc.).
Other natural applications are in
smart lightning systems,
where many small inexpensive
VLB sensors are deployed in several points
of the room to
measure the light intensity and
report the results to the lightning
infrastructures.
Moreover, VLB technologies
can also be employed in security
and anti-introsion systems, as well as
into systems devoted to energy efficiency
\cite{AlAli2017}.}


\subsection{Logistics and industry}

{Wireless technologies, such as
RFID and wireless sensor networks,
are among the important
enabling technologies
for \textit{smart logistics}
\cite{Song2021} and
\textit{industrial IoT (IIoT)} \cite{Liao2018}.
Due to the already mentioned limitations,
RF technologies can be conveniently
replaced by
VLC and VLB ones
in several applications within this domains.}

{A promising use of VLB is that of
indoor localization  and identification
of goods in retailers, shopping malls,
and supermarkets,
as a replacement of RFID or
traditional bar codes.
An example is the VLB technique proposed
in \cite{Wang2016},
aimed at replacing
RFID systems by encoding the information
in a reflective surface,
which is read by using only
ambient illumination (i.e., the sun).
More generally,
the indoor localization capabilities
of VLB can be
applied in several industrial
and logistics fields,
such as real-time
location of assets
in a facility, i.e., tracking of vehicles in industrial sites
or goods in retail shops.
Triangulation-based VLB
techniques based on TOA, TDOA,
or RSSI \cite{Shao2018,Shao2019,Shao2020}
can be used to this aim.
Thanks to the small wavelength of visible light,
these techniques can easily achieve sub-meter
or even centimeter-level accuracy.
In robotics, VLB can be used to provide
precise positioning and
navigation aids in indoor environments.}

{The advantage of VLC and VLB
techniques with respect to RF-based
ones (such as RFID, WiFi, Bluetooth or UWB ones)
is the high localization accuracy attainable and the lack of
the need to install a dedicated
infrastructure, since they can leverage
the existing lighting systems.
Moreover, the presence of a return (UL)
channel allows one to implement
not only client-based localization, but also
server-based one.}

{Finally, VLBC techniques can also be used in
hazardous environments, such as, e.g.,
petrochemical plants, oil ducts, or nuclear plants,
where usage of RF technologies
must be limited of avoided at all.
VLBC techniques can be used for
sensing and monitoring of indoor
infrastructures
such as galleries, ducts, pipelines, etc.}


\section{Future research directions}
\label{sec:future}

Although VLB is a promising paradigm for IoT, there are
some inherent issues to be
dispelled, which deserve further developments  in order to ensure a
widespread use of such a novel technology.
In what follows, we delineate the most
interesting future research directions.

\subsection{Channel modeling for VLB applications}
\label{sec:channel}

{As already pointed out in Section~\ref{sec:channel-modeling},
many works dealing with channel modeling of optical wireless
communications are targeted at the IR region of electromagnetic
spectrum.
However,  there exist significant differences between VLB and IR communications and those results cannot be applied to VLB channel modeling in a straightforward manner.
For instance, an IR source can be approximated as a
monochromatic emitter, while a visible light LED source is inherently wideband.
This fact implies that wavelength-dependency of the source in VLB channel modeling
should be accounted for.
Moreover, in IR communications, the reflectance of materials is typically modeled as a constant.
In contrast, the reflectance of materials in the VLB spectrum should be taken into
consideration due to the wideband nature of VLB links, especially for
the reflection process at the tag.}

{As a matter of fact, a precise characterization of the VLB channel is needed,
by also considering the case of multiple sources and/or hybrid
PD-based and camera-based tags \cite{Teli2022}.
Specifically, advantages and drawbacks of the VLB medium have to
be compared to those of IR media. Physical characteristics of
VLB channels using IM/DD are not fully studied, including path losses
and multipath responses. Another key issue is the characterization  of
natural and artificial ambient VLB noise.}

\subsection{VLBC system throughput}
\label{sec:capacity}

{A technology can be regarded as ``mature'' if
its performance characteristics are well-understood
with well-established design specifications.
In this respect, the {\em throughput} is a measure
of the long-term average rate of a VLBC system,
which  represents a key performance metric
for system designs.
Indoor point-to-point (P2P)
VLC system throughput has been studied in \cite{Matta2019,Jia2020}.
Extension of such works to the VLBC case is not
straightforward, due to further constraints
regarding spatial location
and system geometry.
The results of \cite{Ma2021}
work well for OOK/PAM modulation
and can be applied to
VLBC systems.
System throughput of optical
channels with IM/DD is more complicated due
to some additional constraints that differ from
the conventional electrical or radio systems
\cite{Jiang2016,Wang2013,Xu2017a}.
Specifically, in IM/DD optical systems, the information is modulated as the instantaneous
optical intensity and, therefore,
this peculiarity
places three constraints on the transmitted signal.}

{The first constraint arises from the fact that
the transmitted signal must be non-negative.
Moreover, the eye safety
requirements limit the transmit power that may be used.
Eye safety limitations are
generally expressed in terms of exposure duration at a specific
optical power \cite{Sliney1993}, which
translates to a second {\em average} constraint on the optical power
\cite{Chaaban2022}.
A third {\em peak} constraint also arises
due to safety requirements \cite{Abramovich2011} and,
additionally, in order to avoid saturation of the optical
power (or the device burns).
Evaluation of the system throughput under such three constraints
is challenging even for conventional VLC applications, for which
closed-form expressions are still unknown \cite{Chaaban2022}.}

{In the case of VLBC applications, calculation of the system
throughput is even more complicated due to the additional fact that
the UL channel is {\em double Gaussian}
(i.e., it is the product of two Gaussian random variables)
due to the reflection process performed by the tag.
Therefore, throughput bounds and asymptotics are
essential to understand the ultimate performance limits of
VLBC systems.}

\subsection{Metasurface-based VLB}
\label{sec:metasurfaces}

{The degree of directionality of the source and the tag
significantly impacts on the VLB system performance,
especially when the signal transmitted by the source
is concentrated in a very narrow beam and/or the
tag is characterized by a narrow field-of-view (FOW).
Standard mirrors, such as optical RRs, can support
only specular reflections (i.e., the incident angle and
the reflection angle are identical). Consequently,
mechanical change of their orientation is needed
in order to reflect the beam in a desired direction.}

{An interesting alternative is represented by
the use of gradient {\em metasurfaces} \cite{Tsitsas2017},
which are
synthetic materials composed of
sub-wavelength metallic or dielectric structures
capable of  steering the incident illumination
toward directions not predicted by Snell's law
\cite{Aoni2019,Dolan2021,Ndjiongue2020}.
In optics, they have been used as
reconfigurable intelligent surfaces (RISs)
for the realization of
artificial multichannel
communication systems,
aimed at improving system
performance \cite{Wang2020}
or for energy efficiency
maximization in VLC systems
\cite{Cao2020}.
Moreover, a metasurface
can also be used as a RR, albeit with
much higher efficiency
\cite{Liu2020,Arbabi2017},
or as a VLC modulator
\cite{Sun2019,Kim2020}.}

{Apart from their physical implementation, metasurfaces can be used
in VLB systems in different arrangements. They can replace the RR and/or LCD shutter
modulator, assuring higher efficiency, focusing capabilities, improved speed
and flexibility in implementing more sophisticated modulation/coding schemes.
Another usage is to improve \textit{transmission/reception efficiency}, possibly solving obstruction
problems in LOS links between the source and the tag.
However, there are several issues that can hinder the applicability
of metasurfaces to VLB systems. First, the mathematical modeling of
metasurfaces is generally involved (based on the solution of integral equations),
and simple signal models, useful for system-level design, are lacking.
Second, switching frequencies of current metasurfaces are
inadequate for IoT applications  and
faster switching mechanisms based on
innovative phase transition materials have to be exploited.
Third, existing studies rely on
space-domain design techniques only, i.e.,
the phase  profile of the metasurface is intentionally varied by changing the
state of its sub-wavelength elements
at different  spatial positions on the metasurface.
It would be interesting to also exploit the temporal
dimension of the metasurface, by varying in time the phase
response of its sub-wavelength elements \cite{Dai2018}.}

{Space-time metasurfaces may be used to also overcome
the problem of auto-interference in VLB systems. Indeed,
they allow to control both spatial (propagation direction)
and spectral (frequency distribution) characteristics of the
scattered light, thus allowing to separate at the UL RX
the signal emitted by the source and the signal
backscattered by the tag in the wavelength domain \cite{Zhang2018}.}


\subsection{Multiple access schemes for massive IoT}

{Massive IoT refers to applications that are less latency sensitive and have relatively low throughput requirements, but demand a huge volume of low-cost, low-energy consumption devices
on a network with excellent coverage.
The problem of designing multiple access schemes that are able to deal with the limited
capabilities of the tags is still an open issue for conventional
VLC \cite{Matheus2019} and it
has only recently attracted attention \cite{Pham2020,Yang2022a,Khadr2021,Rishu2022}.
In the literature of VLBC, very few works consider
the problems of supporting multiple tag communications.
A notable exception is RetroI2V \cite{Wang2020a}, where
{\em ad hoc} signaling protocols have been developed to detect
and resolve  collisions in DL/UL of an infrastructure-to-vehicle
communication and networking system.}

{In our opinion, {\em non-orthogonal
multiple-access (NOMA)} schemes \cite{Tse2005} are more suitable for
VBLC in massive IoT applications than their orthogonal counterparts,
since the latter ones may require an unsustainable signaling overhead.
NOMA techniques can broadly be divided
into two categories, namely, power-domain and code-domain
NOMA:
unlike power-domain NOMA, which attains multiplexing in
power domain, code-domain NOMA achieves multiplexing
in code domain.
A power-domain NOMA scheme has been proposed
in \cite{Huang2021} for conventional VLC,
which implements
successive interference
cancellation to remove
interference effects.
The benefits of power-domain
NOMA for VLBC have not been studied yet.
On the other hand, code-domain NOMA schemes
exhibit an inherent robustness against
ambient reflections, i.e., the
interference deriving from
light reflected by other objects
(such as walls or furnitures)
in the ambient, by allowing
easy separation
of the desired signal from
reflections at the RX.
However, code-domain NOMA
could be difficult
to implement with LCD shutter modulators,
due to their limitations in
switch speed \cite{Li2015}.
Use of alternative faster modulators,
like RR-based or metasurface-based ones,
would allow one to use code-domain NOMA schemes,
improving thus system performance.}

\section{Conclusion}
\label{sec:concl}

VLB is a new research field,
where a number of interesting
solutions have been proposed and prototyped,
but many challenges and open problems still exist,
both theoretical and
practical ones.
In this paper, we reviewed
the characteristics and physical-layer treats regarding VLB,
with a focus on IoT indoor applications.
{In particular, VLB are useful whenever
a strong illumination infrastructure
is available and in environments where the lights are always
switched on. It is among the most ``biologically friendly'' and ``green''
techniques.}

{Similar to any wireless communication system, the propagation channel as well as the characteristics of source/tag front-ends dictate the fundamental limits on the physical layer performance
of VLB system. Realistic propagation channel models are therefore of critical importance for VLB system design, performance evaluation and testing.
Moreover, a  particularly interesting field
is the adoption of metasurfaces in VLB, which
could definitely
replace simple RRs and LCD shutters by
allowing increased
flexibility and adaptivity.
Finally, the design of multiple access schemes for VLB channels
might facilitate the realization of the massive IoT vision, according to
which  low cost and low energy sensors, devices, objects, and machines
communicate with each another.}


\bibliographystyle{elsarticle-num-names}
\bibliography{VLC_bib_v5}

\end{document}